# Enantio-specific Detection of Chiral Nano-Samples Using Photo-induced Force

Mohammad Kamandi, Mohammad Albooyeh, Caner Guclu, Mehdi Veysi, Jinwei Zeng, Kumar Wickramasinghe and Filippo Capolino

*Department of Electrical Engineering and Computer Science, University of California, Irvine, California 92697, USA*

We propose a novel high resolution microscopy technique for enantio-specific detection of chiral samples down to sub-100 nm size, based on force measurement. We delve into the differential photo-induced optical force $\Delta F$ exerted on an achiral probe in the vicinity of a chiral sample, when left and right circularly polarized beams separately excite the sample-probe interactive system. We analytically prove that $\Delta F$ is entangled with the enantiomer type of the sample enabling enantio-specific detection of chiral inclusions. Moreover, we demonstrate that $\Delta F$ is linearly dependent on both the chiral response of the sample and the electric response of the tip and is inversely related to the quartic power of probe-sample distance. We provide physical insight into the transfer of optical activity from the chiral sample to the achiral tip based on a rigorous analytical approach. We support our theoretical achievements by several numerical examples, highlighting the potential application of the derived analytic properties. Lastly, we demonstrate the sensitivity of our method to enantio-specify nanoscale chiral samples with chirality parameter in the order of 0.01, and discuss how this could be further improved.

## 1. INTRODUCTION

Chirality is of supreme importance in life sciences. This significance originates from the fact that the fundamental building blocks of life, i.e., proteins and nucleic acids are built of chiral amino acids and chiral sugar, respectively [1]. More importantly, in pharmaceutical applications, the enzymes and receptors of human body can constructively react to drugs with only proper enantiomers as known as the Fischer's "lock and key" principle [2]. Therefore, if a drug is racemic, i.e., contains both mirror image enantiomers, namely R and S enantiomers, then it might result in detrimental effects beside the desired ones. As a result, based on the importance of chirality in chemical and pharmaceutical sciences, detection and characterization of chiral particles (materials) are fundamentally critical issues. In order to resolve these issues, spectroscopy techniques based on optical rotation (OR), circular dichroism (CD), and Raman optical activity (ROA) have been proposed [3–6]. In these chiroptical techniques, the scattered (refracted or absorbed) light from the sample is measured to detect the chirality. In particular, by measuring the difference in absorbance of the right and left hand circularly polarized (CP) light in CD, not only the chirality but also the primary and secondary structure of sample (i.e. proteins) can be determined [7–9]. However, down to molecular scales, due to the very weak interaction of light with chiral nanoparticles, the spectroscopic techniques for chirality detection encounter major challenges. For instance, the background scattering noise would play a notable prohibiting role in the chirality detection by using these techniques. Although many studies have been performed to resolve the background noise problem [10,11], still a considerable amount of the material is required for the detection process. These challenges call for suitable techniques for the development of chirality detection at nanoscales.

Recently, optical force, as a powerful tool for a wide variety of applications from trapping [12–15] and manipulating of nanoparticles [16–18] to optical cooling [19–21] and imaging the electromagnetic fields [22], has been brought up for separation or manipulation of enantiomers [23]. Indeed, in contrast to the long-term common belief that particles are always pushed by light, in [24–27] lateral and pulling optical forces were discovered for chiral particles. Therefore, by taking the advantage of the discriminatory behavior of enantiomers, due to their optical activity, Tkachenko et al. in [28] have shown how to sort enantiomers in fluidic environments for micro-meter sized particles. In [29], separating of enantiomers has been expanded to nano-meter sized particles by the same concept using a plasmonic tweezer that rely on enhanced near-field gradients. However, optical force has not been exploited for characterization and enantio-detection of chiral particles.

The importance of detecting chirality at nanoscale in one hand and the lack of high resolution reliable measurement techniques on the other hand urges us to explore possible techniques for detection of chirality of nanoscale particles. Very recently, combining the advantages of atomic force microscopy (AFM) [30] and optical illumination in the so-called photo-induced force microscopy (PiFM) [31–33], has provided the possibility of probing linear [34] and non-linear [35] optical characteristics at the nanoscale with higher accuracies compared to scattering measurement techniques. However, to the best of our knowledge, this encouraging technique which is highly appropriate and accurate to extract optical properties of materials (down to nanoscales) has never been opted to detect the materials' chirality.

In this work, we introduce for the first time the use of PiFM for enantiomeric detection of nanoscale chiral samples. We propose to probe the difference between the force exerted on the AFM tip in the vicinity of a sample for CP wave illuminations with opposite handedness, i.e., right and left hand. Though material chirality is a weak effect in the light-matter interaction compared to the electric property, we take its effect into account by using the dipolar approximation and considering both electric and magnetic dipole moments. This way, we provide analytical arguments which prove that for chiral samples, the proposed differential photo-induced force is nonzero whereas for achiral ones it vanishes. More importantly we demonstrate that for a specific enantiomer, the differential force is equal with opposite sign to its mirror image which paves the way toward revealing the enantiomer type for a chiral sample at nanoscale. Notice that conventional chiroptical methods such as CD for chirality detection are based on measurements related to averaged far-field scattering from the sample and require a substantial amount of a chiral material. Instead, our proposed technique is capable of enantio-specifically detect the chirality of samples down to sub-100 nm size with a chirality parameter of $\kappa = 0.04$. Future developments may even lead to the detection of weaker chirality.

The main goal of this work is to provide a method for detecting material chirality at nanoscale in the broad sense. The considered material samples could be made for example by a concentration of chiral molecules or of an engineered chiral nanoparticle.

The paper is structured as follows. In Section II, we outline the physical principle of PiFM operation by providing the model for tip-sample system and general formulation for the exerted force on the tip. In Sec. III we prove the concept of using PiFM as a chirality sensor and provide a simple formulation to predict the differential photo-induced force on the tip and examine its accuracy by several examples. Then, in section IV, we demonstrate the physics behind the tip-sample interactive system. We exhibit the potential dynamic range of our proposed method in detection of chirality at nanoscale in section V. At the end, we conclude the paper with some remarks.

## 2. OBJECTIVE AND GENERAL PHYSICAL PRINCIPLE

Fig. 1(a) shows the photo-induced force measurement set-up using the PiFM that is investigated in this paper. The system of nanoscale sample and the microscope tip are illuminated by an incident light from the bottom side. Incident light induces polarization currents on both the sample and tip. The sample and tip are located at the near-fields of their re-scattered fields so that they exert a notable amount of force on each other in the normal direction (the $z$-direction in Fig.1) due to the gradients of their near-fields. The goal of this paper is to show that it is possible to discriminate with respect to the handedness of the chiral sample particle based on applying CP light, with separately left and right polarization and measuring the force exerted to the tip for both polarizations and obtain the force difference. In order to exemplify the fundamental physical principle of operation, we consider the sample and tip radius to be optically small. Therefore, we may model both the sample and tip as two particles illuminated by an external electromagnetic field with wave vector k [Fig. 1(b)]. The sample is characterized by a bianisotropic response (chirality is a special case) that provides both electric $\mathbf{p}_s$ and magnetic $\mathbf{m}_s$ dipolar moments (where subscript "s" represents "sample"). Indeed, the consideration of magnetic dipole moment is necessary in the analysis of the electromagnetic response of chiral particles since under electromagnetic illumination, a sample composed of chiral inclusions exhibits optical activity. Under the dipolar approximation, valid for subwavelength particles, such as molecule concentrations or engineered "meta-atoms", the optical activity is well described by the introduction of the electric dipole-to-magnetic dipole (magnetoelectric) polarizability [36]. Notice, although this magnetoelectric polarizability is weak compared to that associated to the electric dipole polarizability, neglecting it in the analysis of chiral particles will result in an incorrect prediction of the electromagnetic response of the sample. Based on the above discussion, by using the polarizability tensor of such a chiral particle, the electric and magnetic dipole moments read [37–39]

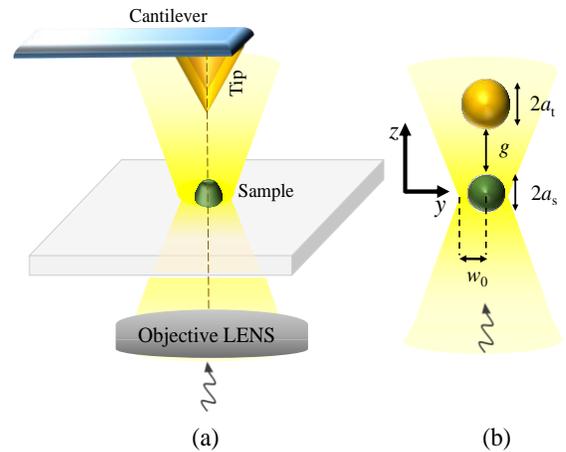

FIG. 1. (a) Schematic of photo-induced force microscopy capable of detecting the chirality of a sample based on differential force measurement. (b) Simplified model where the tip and sample are represented by nanospheres with electric (and perhaps magnetic) dipole moment(s).

$$\begin{bmatrix} \mathbf{p}_s \\ \mathbf{m}_s \end{bmatrix} = \begin{bmatrix} \underline{\alpha}_s^{ee} & \underline{\alpha}_s^{em} \\ \underline{\alpha}_s^{me} & \underline{\alpha}_s^{mm} \end{bmatrix} \begin{bmatrix} \mathbf{E}^{loc}(\mathbf{r}_s) \\ \mathbf{H}^{loc}(\mathbf{r}_s) \end{bmatrix} \quad (1)$$

here $\mathbf{E}^{loc}$ and $\mathbf{H}^{loc}$ are, respectively, the phasors of the local electric and magnetic field at the sample location $\mathbf{r}_s$. Moreover, $\underline{\alpha}^{ee}$, $\underline{\alpha}^{mm}$, $\underline{\alpha}^{em}$, and $\underline{\alpha}^{me}$ are, respectively, electric, magnetic, magnetoelectric and electromagnetic polarizability tensors of the sample particle. The last two polarizability tensors are also called bianisotropic parameters which relate the local electric (magnetic) field to the magnetic (electric) dipole moment. In general, there are four classes of bianisotropic particles, i.e., omega, chiral, Tellegen, and "moving", which any dipole particle can be reduced to. The first two classes are reciprocal while the second two are nonreciprocal. Reciprocity implies $\underline{\alpha}^{me} = -\left(\underline{\alpha}^{em}\right)^T$, where superscript "T" denotes the tensor transpose. In addition, for a pure chiral sample particle, the bianisotropic tensors are diagonal, hence one has $\alpha_{jj}^{me} = -\left(\alpha_{jj}^{em}\right)$, where $j=x,y,z$. Under the assumption of negligible sample losses $\alpha_{jj}^{me}$ is purely imaginary [38].

In the following, we assume that the chiral sample has a spherical shape and has isotropic response (as in the case it contains an amorphous arrangement of many chiral molecules). The sample's isotropic response implies it has equal polarizability components in all spatial directions. Under this assumption, the electric and magnetic polarizability tensors reduce to $\underline{\alpha}^{ee} = \alpha^{ee}\underline{\mathbf{I}}$ and $\underline{\alpha}^{mm} = \alpha^{mm}\underline{\mathbf{I}}$, respectively, where $\underline{\mathbf{I}}$ is the identity tensor. Moreover, for a chiral isotropic particle $\underline{\alpha}^{em} = \alpha^{em}\underline{\mathbf{I}}$ and $\underline{\alpha}^{me} = -\left(\underline{\alpha}^{em}\right)^T = -\alpha^{em}\underline{\mathbf{I}}$ (where superscript "T" denotes the tensor transpose) would be the magnetoelectric and electromagnetic polarizability tensors, respectively. The expressions for polarizabilities of an isotropic spherical chiral particle in terms of its permittivity, permeability, and chirality parameter, obtained based on Mie scattering theory, are given in Appendix A. We assume the tip is achiral and model its response to an electromagnetic wave by electric and magnetic dipole moments

$$\mathbf{p}_t = \underline{\alpha}_t^{ee} \cdot \mathbf{E}^{loc}(\mathbf{r}_t), \quad \mathbf{m}_t = \underline{\alpha}_t^{mm} \cdot \mathbf{H}^{loc}(\mathbf{r}_t). \quad (2)$$

Notice, the local electromagnetic field is considered at the tip location $\mathbf{r}_t$, where subscript "t" represents "tip". Also, the local field at the tip position is the contribution of the external incident and the scattered fields of the sample, which is modeled as a dipolar system with the electric and magnetic moments as defined in Eq. (1). Considering this

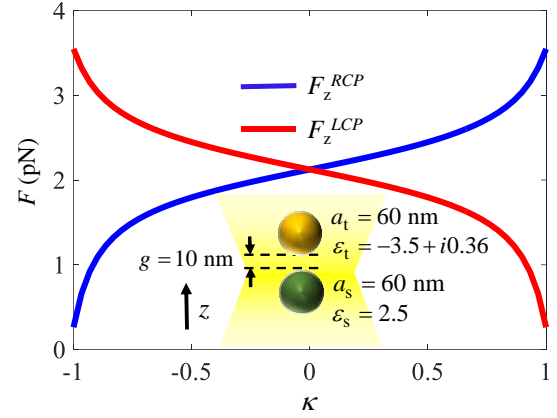

FIG. 2. The effect of sample chirality on the induced force on the tip for two incident scenarios of RCP and LCP light.

model, the general expression of the time-averaged optical force exerted on the tip is given by [27]

$$\langle \mathbf{F} \rangle = \frac{1}{2}\text{Re}\left[ \mathbf{p}_t \cdot \left(\nabla \mathbf{E}^{loc}(\mathbf{r}_t)\right)^* + \mathbf{m}_t \cdot \left(\nabla \mathbf{H}^{loc}(\mathbf{r}_t)\right)^* - \frac{ck^4}{6\pi}\left(\mathbf{p}_t \times \mathbf{m}_t^*\right) \right], \quad (3)$$

where the asterisk denotes complex conjugation, $c$ is the speed of light and $k$ is the wavenumber of host medium. Moreover, $\nabla\mathbf{E}$ and $\nabla\mathbf{H}$ are two tensors of the second rank (the gradient of a vector is defined in the Appendix B). In Eq. (3), the first and second terms represent the force acting on the corresponding electric and magnetic dipoles of the tip, respectively, while the third term represents the force due to the interaction between both induced electric and magnetics dipoles of the tip. We calculate the exerted force on the tip in the rest of the paper by using the above formalism. In this paper every field is monochromatic and the time convention $\exp(-i\omega t)$ is implicitly assumed and suppressed.

## 3. PIFM AS CHIRALILTY SENSOR AT NANOSCALE

In this section, we illustrate the potential of PiFM in distinguishing enantiomer type of nanoscale samples. Based on the force formulation discussed in section II, i.e., by using dipole approximation limit and Mie scattering theory to calculate the polarizability of spherical chiral sample nanospheres, we examine the force induced on the achiral tip for different scenarios. As was discussed, we illuminate the tip-sample system with an incident light from the bottom side (see Fig.1). We selectively apply CP light with both handedness, i.e. right and left hand CP (RCP and LCP) light. The induced polarizations on an *achiral* sample would be the same for the RCP and LCP beams. This results in identical re-scattered fields for an achiral sample when excited with the proposed RCP or LCP. Therefore, the exerted force on the tip alone (along the propagation *z*-direction) is the same for opposite sense of handedness of the incident light. In contrast, by virtue of its optical activity, the induced polarizations $\mathbf{p}_s$ and $\mathbf{m}_s$ on a chiral sample are not equal for opposite sense of incident light handedness, and hence, the re-scattered near-fields are different. Therefore, for a chiral sample, the exerted force on the tip would be different for opposite sense of handedness of the incident light. To verify this, we consider an exemplary case when the sample and plasmonics tip are both considered to have equal radii $a_s = a_t = 60$ nm which are at a distance to form a particle-to-particle gap of $g = 10$ nm (see Fig. 3). Moreover, without loss of generality, the relative permittivity of the sample is assumed to be $\varepsilon_s = 2.5$ whereas the plasmonics tip is assumed to have a relative permittivity $\varepsilon_t = -3.5 + i0.35$ (this choice will be discussed later). The incident light is assumed to be CP Gaussian beam propagating along the *z*-direction with wavelength $\lambda = 504$ nm with 1mW power. The waist of the Gaussian beam (see Fig.1) is set to $w_0 = 0.7\lambda$ and is positioned at the *z*=0 plane, where the sample is located, because of the higher strength of the field at the beam waist along the beam axis. The sample is made of a sphere composed of chiral material described by the chirality parameter $\kappa$ that here varies from -1 to 1 (see Appendix A). By using the introduced analytical formalism, the force expression (3), we have calculated the *z*-component of the induced force on the tip $F_z^{RCP}$ and $F_z^{LCP}$ versus chirality parameter $\kappa$, for the two excitation scenarios with RCP and LCP beams. The results in Fig. 2 show that there is no force difference between two cases of RCP and LCP incidences for $\kappa = 0$. However, as we increase the amplitude of the chirality parameter $\kappa$ of the sample, the differences between induced forces on the tip becomes more obvious, in the order of piconewtons. Moreover, if we define the *differential* photo-induced force as

$$\Delta F = F_z^{RCP} - F_z^{LCP}. \quad (4)$$

We note that for a pair of enantiomer $\Delta F$ is equal in amplitude but opposite in sign (we recall that the chirality parameters of an enantiomer and its mirror image have equal amplitude with opposite sign). This example clearly demonstrates that by measuring differential photo-induced force, we can differentiate between chiral enantiomers.

To unravel the physical principle behind this interesting discriminatory behavior of chiral particles, we note that the local field acting on the sample to be used in Eq. (1), provided by both the incident field and the near-field generated by the tip, is found by

$$\begin{aligned}\mathbf{E}^{loc}(\mathbf{r}_s) &= \mathbf{E}^{inc}(\mathbf{r}_s) + \underline{\mathbf{G}}^{EP}(\mathbf{r}_s,\mathbf{r}_t) \cdot \mathbf{p}_t, \\ \mathbf{H}^{loc}(\mathbf{r}_s) &= \mathbf{H}^{inc}(\mathbf{r}_s) + \underline{\mathbf{G}}^{HP}(\mathbf{r}_s,\mathbf{r}_t) \cdot \mathbf{p}_t.\end{aligned} \quad (5)$$

Here $\underline{\mathbf{G}}^{EP}$ and $\underline{\mathbf{G}}^{HP}$ are the dyadic Green's functions that provide the electric and magnetic fields, respectively, generated by an electric dipole [39]. As was discussed earlier, we investigate the case of an *isotropic achiral tip*, whose dipole moments $\mathbf{p}_t$ and $\mathbf{m}_t$ are given by Eq. (2). Moreover, in Eq. (5) we have assumed that the magnetic dipole moment $\mathbf{m}_t$ of the tip is negligible which we clarify later to be an acceptable approximation for a tip made of plasmonic material (note that the results of Fig. 2 are obtained considering all dipole terms in Eq. (3) including a very small $\mathbf{m}_t$, however, in deriving an approximate formula for the force we have neglected the magnetic dipole moment $\mathbf{m}_t$ of the tip). The local electric field at the tip is provided by both the incident field and the near-field scattered by the sample that is assumed to possess both electric $\mathbf{p}_s$ and magnetic $\mathbf{m}_s$ dipole moments:

$$\mathbf{E}^{loc}(\mathbf{r}_t) = \mathbf{E}^{inc}(\mathbf{r}_t) + \underline{\mathbf{G}}^{EP}(\mathbf{r}_t,\mathbf{r}_s) \cdot \mathbf{p}_s + \underline{\mathbf{G}}^{EM}(\mathbf{r}_t,\mathbf{r}_s) \cdot \mathbf{m}_s. \quad (6)$$

Here $\underline{\mathbf{G}}^{EM}$ is the dyadic Green's function that provides the electric field generated by a magnetic dipole. The expression for the time-averaged optical force exerted on the achiral tip (neglecting the effect of magnetic dipole i.e., $\mathbf{m}_t = 0$) reads

$$\langle \mathbf{F} \rangle = \frac{1}{2}\mathrm{Re}\left[\mathbf{p}_t \cdot \left(\nabla \mathbf{E}^{loc}(\mathbf{r}_t)\right)^*\right]. \quad (7)$$

Calculations are done by using all the dynamic terms in the Green's function. Next, neglecting the field's phase difference between the tip and sample, due to their subwavelength distance, it can be shown (see Appendix B for more details) that the difference between the forces exerted on the tip for two CP plane waves with opposite handedness reads

$$\Delta F \approx -\frac{3|\mathbf{E}_0|^2}{4\pi\sqrt{\varepsilon_0\mu_0}d^4}\mathrm{Im}\left\{\alpha_t^{ee}\alpha_s^{em}\right\}. \quad (8)$$

In deriving this approximate but physically insightful formula we have assumed the axial Gaussian beam field to be approximated with a CP plane wave with electric field magnitude $|\mathbf{E}_0|$. Furthermore, $\varepsilon_0$ and $\mu_0$ are, respectively, the free space permittivity and permeability, and $d$ is the center-to-center distance between tip and sample. Eq. (8) is a striking result since it clearly demonstrates that in the absence of the magnetoelectric polarizability (i.e., $\alpha_s^{em} = 0$) $\Delta F$ would be zero, while for a chiral sample it is not the case. Therefore, we can distinguish between chiral and achiral samples by observing $\Delta F$. More importantly, as is known, the quasi-static approximation for magnetoelectric polarizability $\alpha^{em}$ of a chiral sphere with material parameters $\varepsilon_S$, $\mu_S$ and $\kappa$ is [40]

$$\alpha_s^{em} = 12i\pi a^3 \sqrt{\varepsilon_0 \mu_0} \frac{\kappa}{(\varepsilon_s + 2)(\mu_s + 2) - \kappa^2}, \quad (9)$$

where $a$ is the radius of the sphere, and it is considered that $\mu_s = 1$. From Eq. (9) the sign of magnetoelectric polarizability $\alpha_s^{em}$ is dictated by the sign of chirality parameter of the sample $\kappa$. Therefore, Eq. (8) expresses that $\Delta F$ is equal in amplitude but opposite in sign for a pair of enantiomer, which affirms that our proposed method is capable of distinguishing enantiomer type. It is also worth mentioning that based on Eq. (8) and (9), $\Delta F$ is linearly proportional to chirality parameter $\kappa$ of the sample (since $\kappa^2$ in the denominator of Eq. (9) is much smaller than the first term for natural materials) within the quasi-static approximation. Additionally, Eq. (8) states that $\Delta F$ is linearly dependent to the electric polarizability of the tip, emphasizing the importance of the tip material and its geometry. In our example we have assumed to have a plasmonic tip to boost its electric dipolar response and hence increasing the electric polarizability contribution. Furthermore, $\Delta F \propto 1/d^4$ (where $d$ is the center-to-center distance between the tip and the sample) which illustrates the sensitivity of the probing force to the distance between the tip and the sample.

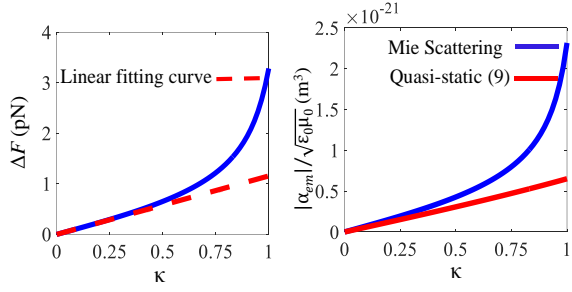

FIG.3. a) $\Delta F$ versus chirality parameter $\kappa$ of the sample. Tip and sample spheres' radii are $a_s = a_t = 60$ nm and $\varepsilon_t = -3.5 + i0.35$ and $\varepsilon_s = 2.5$, and gap is $g = 10$ nm. For completeness we show a linear fitting curve for $\Delta F$ b) Normalized magnetoelectric polarizability of the sample using Mie formula given in Appendix A and the quasi-static approximation Eq. (9).

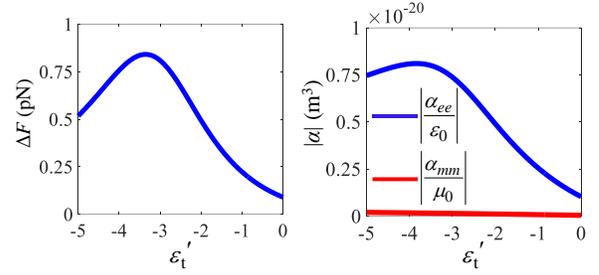

FIG. 4. Differential force $\Delta F$ and normalized electric and magnetic polarizability of the tip as a function of the real part of the tip relative permittivity. A plasmonics tip provides a stronger electric polarizability and hence a stronger force.

We have shown the main physical principle and we now dig into Eq. (8) to demonstrate how our analytical formalism boast itself in predicting the general trends of differential photo-induced force. To that end, we first emphasize on the dependence of the differential force on the chirality parameter $\kappa$. We assume equal radii for the tip and sample i.e., $a_s = a_t = 60$ nm with the gap $g = 10$ nm between the tip and sample. Furthermore, we consider the relative permittivity of the tip and sample to be $\varepsilon_t = -3.5 + i0.35$ and $\varepsilon_s = 2.5$, respectively, as was done for the result in Fig. 2. We then sweep over the chirality parameter of the tip from 0 to 1. By applying Eq. (3), we plot the differential induced forces $\Delta F$ on the tip for the two different incidence scenarios of RCP and LCP beams (solid blue line in Fig. 3(a)). As it is clear, $\Delta F$ shows linear dependence on $\kappa$ when its chirality parameter is smaller than 0.5. For $\kappa > 0.5$ the differential force $\Delta F$ diverges from its linear fitting curve, the red dashed line in Fig. 3(a) which presents the asymptotic linear behavior of the differential force for small chirality parameter $\kappa$. To provide an insight for the reason of this behavior, we have plotted the normalized (to $1/c$) magnetoelectric polarizability of the sample as a function of chirality parameter [see Fig. 3(b)], calculated with two methods; i.e., the exact calculation using Mie coefficients (see Appendix A) and the approximate quasi-static formulation in Eq. (9). As it can be seen, for $\kappa < 0.5$ the results of both exact and approximate methods absolutely match whereas for higher values of $\kappa$ the quasi-static approximation diverges from the accurate Mie coefficient result. Thus, $\alpha^{em}$ and hence $\Delta F$ are assumed to be linearly dependent on $\kappa$ for $\kappa < 0.5$. Indeed the direct comparison between $\Delta F$ plotted in Fig 3(a) and the magnetoelectric polarizability $\alpha^{em}$ of the sample calculated through the exact Mie coefficient (in Appendix A) shown in Fig 3 (b) is decoding their linear dependence. This result is exactly what we previously stated based on Eq. (8).

As one can infer from the approximate formula (8), the next crucial parameter in determining differential induced force is the tip electric dipolar polarizability $\alpha_t^{ee}$. In Fig. 4 we provide $\Delta F$ and normalized electric and magnetic polarizabilities as functions of real part of the tip relative permittivity given by $\varepsilon_t = \varepsilon_t' + i0.35$ assuming $a_s = 60$ nm. As shown in Fig. 4(a), $\Delta F$ peaks around $\varepsilon_t' = -3.5$ which corresponds to the electric resonance of the tip which is demonstrated in Fig. 4(b). From a comparison between Figs.

4(a) and 4(b) we conclude that $\Delta F$ and $\alpha_t^{ee}$ reveal similar behaviors as we sweep the real part of permittivity which once more proves the potential of using Eq. (8) to state that $\Delta F$ is linearly dependent on the tip electric polarizability $\alpha_t^{ee}$. (Force calculation is done via Eq. (3) using the dynamic Green's function.) Notice that the magnetic polarizability is much smaller than its electric counterpart, emphasizing the negligible contribution of the second and third terms of Eq. (3). As a result of the above discussion, we conclude that the choice of material parameter for the tip is of quite high importance because the dynamic range of experimental measurement setup is limited by noise and the force value may be small depending on particle size, laser power, and materials. As we will discuss in the next section and as demonstrated in Eq. (8), a tip with strong electric polarizability enabled by plasmonic material enhances the measured photo-induced force response of a chiral sample and hence the differential force. Our goal is to enantio-specify samples with nanoscale size (microscopic chirality identification), despite the weak chiral response of their matter constituent, by using plasmonic material for the tip. It should be noted that the tip shape could also be engineered to provide even higher electric polarizability compared to a simple plasmonic sphere [11,29,41-46] (e.g. a triangular prism or truncated tetrahedron). However, considering the fact that our aim in this paper is to introduce the technique and to prove its capability, we assumed a plasmonic sphere (simplest shape) tip in our analysis. Nevertheless, in future study involving also experimental verification, the tip shape could be engineered to further enhance the ability in the detection of observable response of the chiral sample. Note that however shorter distances $d$ may provide a much stronger force but we cannot show it here since the dipolar representation of tip and sample loses validity for smaller distances. Indeed, the last central parameter in the enantio-specific detection of chiral nano-samples is the tip-sample distance as illustrated in the differential force approximate expression Eq. (8). In Fig. 5, we have depicted $\Delta F$ in logarithmic scale, evaluated by Eq. (3), as a function of the tip-sample distance from 75 nm to 175 nm along with a $d^{-4}$ dependent function for a tip-sample system with parameters as follows: $a_s = a_t = 25$ nm, $\varepsilon_t = -3.5 + 0.35i$, $\varepsilon_s = 2.5$ and $\kappa = 0.6$. As shown, there is a good agreement between the force and $d^{-4}$ function which was again predicted by Eq. (8).

So far, we have analytically proved that by using PiFM we not only detect chiral samples by using achiral probes, but also specifically determine the enantiomer type of a chiral sample. We have supported our analytical findings by numerical calculations. In the next section, by using analytical formulations we present the *effective* polarizability model for an achiral tip when it is closely positioned near a chiral sample, hence, give physical insight to explain the phenomena happened in the aforementioned discussion.

## 4. OPTICAL ACTIVITY TRANSFER FROM CHIRAL SAMPLE TO ACHIRAL TIP

We provide here a simple formulation to predict the differential photo-induced force on the tip in the vicinity of a chiral sample. In this section we aim at delivering an equivalent representation of electric dipole moment which presents deep physical insight into a unique phenomenon: i.e., the transfer of optical activity from a chiral particle to an achiral one in its vicinity. In this equivalent representation, we rewrite the electric dipole moment of the tip as

$$\mathbf{p}_t = \hat{\alpha}_t^{ee}\mathbf{E}^{\text{inc}} + \hat{\alpha}_t^{em}\mathbf{H}^{\text{inc}}. \tag{10}$$

in which $\hat{\alpha}_t^{ee}$ and $\hat{\alpha}_t^{em}$ are the *effective* electric and magnetoelectric polarizabilities of the tip when it is in the close vicinity of the chiral particle. As it is clear, in this representation, instead of local electric and magnetic fields, we use incident fields. In other words, we include the presence of the chiral particle near-field by modifying the polarizability of the tip. As shown in the Appendix C, effective electric and magnetoelectric polarizabilites are given by

$$\hat{\alpha}_t^{ee} = \alpha_t^{ee}\frac{1-\alpha_s^{ee}G}{1-\alpha_t^{ee}\alpha_s^{ee}G^2}, \tag{11}$$

$$\hat{\alpha}_t^{em} = -\alpha_t^{ee}\frac{\alpha_s^{em}G}{1-\alpha_t^{ee}\alpha_s^{ee}G^2}, \tag{12}$$

in which $G$ is defined in (B4). It is very important to notice that although the tip is not chiral and is modeled by a simple electric dipole, the impact of the chiral sample makes the tip to effectively act as a chiral particle as in Eq. (10). That is to say, a chiral sample can induce chirality to an achiral particle if placed in vicinity of it which can be utilized in two ways:

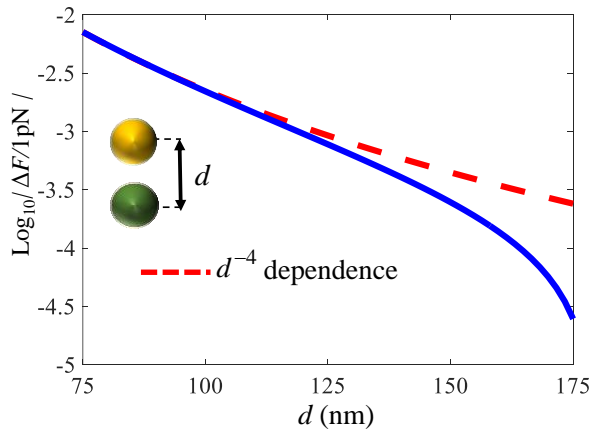

FIG. 5. $\Delta F$ in logarithmic scale versus the distance between the tip and sample (solid blue line) along with a $d^{-4}$ dependent function (red dashed line).

1) creating artificial optically active particles using achiral materials, and 2) using plasmonic particles as a "reporter" (or equivalently an antenna) for chirality detection. Here we engineer a gold nanoparticle to be optically active by placing it in the vicinity of a chiral particle. Thanks to the strong electric resonance of plasmonic particles, the optical activity of the tip-sample interactive system will be enhanced compared to a single chiral particle. The advantage of creating a chirality reporter using a plasmonic particle can be understood by observing that the chiral response of most biomolecules peaks in UV regime, however, in the vicinity of a plasmonic particle, this signature can be brought to plasmonic band, making the measurement easier in the visible region. Although this phenomena has been already reported in [46–50], here we have provided a rigorous analytic representation of it which allows to design proper antennas in order to maximize magnetoelectric coupling and hence improving chirality detection. Using this equivalent approach, we can show that the differential induced force on the tip at the close vicinity of the sample reads (see Appendix C)

$$\Delta F = F_z^{RCP} - F_z^{LCP} = \\ -\frac{|\mathbf{E}_0|^2}{|\alpha_t^{ee}|^2} \text{Im}\left\{\alpha_t^{ee}\left(\hat{\alpha}_t^{ee}\frac{\partial \hat{\alpha}_t^{em*}}{\partial z} - \hat{\alpha}_t^{em}\frac{\partial \hat{\alpha}_t^{ee*}}{\partial z}\right)\right\}\Bigg|_{z=z_t},$$
(13)

in which $\partial/\partial z$ represents partial derivative with respect to the position on the $z$-axis. This formula further emphasizes the dependence of $\Delta F$ on the effective parameters of the tip, i.e., effective electric and magnetoelectric polarizabilities.

We conclude this section by stating that our analytical formalism allows us to suggest another approach to maximize the signature of optical activity of nanoparticles by using properly engineered nano-antennas which enhance both electric and magnetic fields rather than only electric field [51] which is what we have shown here. Notice, the magnetic dipole of the antenna must be properly oriented in order to effectively couple to the chirality of sample particle. That is, if the antenna can be represented by dipole moments, its electric and magnetic moments must be collinear which implies a chiral response. To obtain such strong coupling between the antenna and the sample particle, and hence, a stronger optical response of the tip, we suggest to take the advantage of a proper combination of structured light excitation and antenna near-field response which can be a subject for further studies. For instance, one may use a chiral antenna [52,53] with CP excitation. Alternatively, one may use an achiral antenna structure with both electric and magnetic responses and use a combination of the structured light as the excitation scheme to control the coupling between chiral samples and the tip.

In the next section, we provide a measure to clarify the dynamic range of our approach in probing chirality of optically small nano-samples down to sub-100 nm sizes.

## 5. POTENTIAL OF THE APPROACH IN DETECTING NANOSCALE CHIRAL INCLUSIONS

So far, we have demonstrated the possibility of detecting chirality and the enantiomer-type of nanoscale samples using photo-induced force. However, we have not yet discussed the resolution of our proposed method i.e., the minimum size of the specimen which results in a detectable $\Delta F$ and how misalignment of the tip affects the chirality detection. To investigate this first issue, in Fig. 6 we make a color map showing $\Delta F$ in the logarithmic scale versus chirality parameter $\kappa$ (again in logarithmic scale) of the sample and radius of the sample $a_s$. We have assumed that the radius of the tip is $a_t = 60$ nm and the gap is $g = 10$ nm and the sample is positioned at $z=0$ plane on the axis of the excitation beams at the minimum beam waist. As before, we assume relative permittivities of the tip and sample to be the same as Fig. 2. The chosen chirality parameter range binds the chirality parameter of common chiral specimens including DNA-assembled nanostructures and composite nanomaterials [29]. We have marked the 0.1pN force boundary with a black solid line. For the region above this black line, $\Delta F$ is greater than 0.1pN. This number is chosen based on [54] to be the instrument general sensitivity. It can be seen, the smallest detectable chirality parameter value $\kappa$ for the maximum studied sample radius $a_s = 70$ nm is 0.04 (shown with $\log_{10}|\kappa| = -1.39$).

It is worth noting that a chirality parameter of the order of $\kappa \sim 10^{-2}$ which corresponds to a specific rotation[1] of $[\alpha]_D \sim 100000°$ is still a giant value compared to that obtained with chiral molecules such as Glucose (C6H12O6), Carvone (C10H14O), Testosterone (C19H28O2), etc., with a chirality parameter of the order of $\kappa \sim 10^{-6}$ [55–58] (which corresponds to a specific rotation in the order of $[\alpha]_D \sim (100-200)°$). This giant chirality is only observed in a few molecules and compounds such as helicene or Norbornenone with specific rotation angles of the order of $[\alpha]_D \sim 100000°$ [59,60]. However, we emphasize that our technique, in contrast with conventional chiroptical techniques, is capable of detecting such molecules when the size of a sample is in the order of ~100nm. The detection of chiral samples with smaller chirality parameter requires some extra considerations, like an increase of the incident power which is possible in cryogenic conditions [61] or in liquids to maintain the condition of gold coated tip [62–64] or using nanotips with stronger electric polarizability, or even explore nanotips able to express magnetic response.

---

[1] Specific rotation is defined as the optical rotation in degrees per decimeter divided by the density of optically active material in grams per cubic centimeter [1].

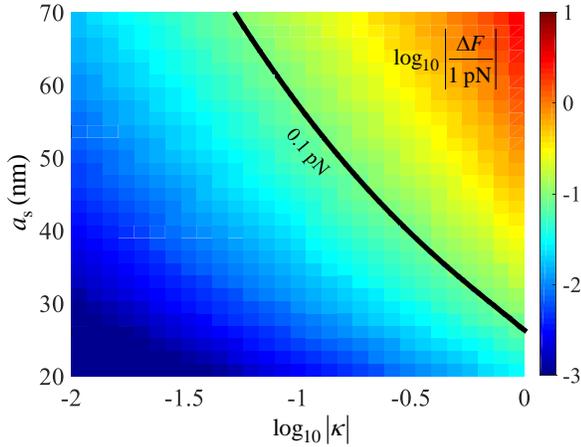

FIG. 6. Illustration of the sensitivity of our proposed technique. $\log_{10}|\Delta F/1\text{pN}|$ is shown versus chirality parameter and radius of the sample.

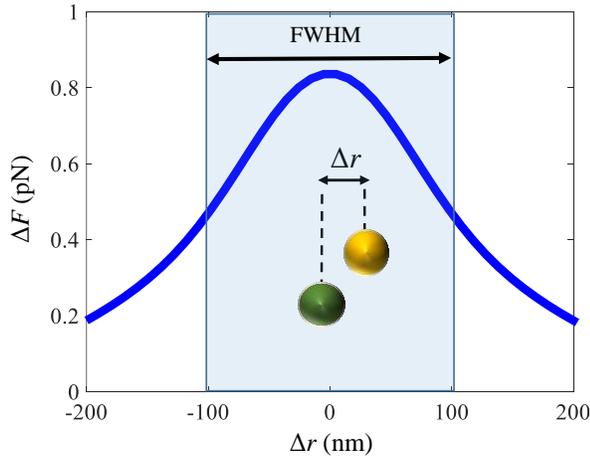

FIG. 7. $\Delta F$ versus lateral displacement of the tip versus sample. As the tip moves laterally the differential force decreases.

Notice, on the other extreme side of the studied ranges of parameters, the smallest detectable radius for a sample with maximum chirality parameter $|\kappa|=1$ (shown with $\log_{10}|\kappa|=0$) is 25 nm.

Next, we investigate the transverse resolution of the proposed PiFM technique by observing the detectability range of the sample by assuming a controlled misalignment of the tip. It is expected that this lateral misalignment will decrease the probed differential force. In the following we estimate the full width at half maximum (FWHM) of the differential force. This is defined as the range of lateral distances between tip and sample in which the differential force is higher than the half of its maximum value (i.e., when the lateral distance between the tip and sample is zero) and shown in Fig. 7. In this figure, the differential force is depicted versus the lateral distance between the tip and sample assuming the system parameters as in Fig. 2 and $\kappa=0.6$. As it is expected, when the tip moves laterally (in either direction), the differential z-directed force decreases and FWHM of the differential force is 200 nm. This example shows also how far the tip could be to be able to make an enantio-specific detection of chiral nano-samples. One may refer to section VI for more analyses on different parameters influencing the probed differential force such as the relative positioning of the tip-sample interactive system and the excitation beam, the tip-sample gap, the tip radius, etc.

## 6. FORCE DEPENDENCE ON PHYSICAL PARAMETERS

In this section we investigate the dependence of the induced force on the tip on its radius and on the tip-sample relative displacement with respect to the excitation beam. Incident light is assumed to be a CP Gaussian beam propagating along the positive z-direction with wavelength $\lambda=504$ nm and 1mW power. The minimum waist of the Gaussian beam (see Fig.1) is positioned at the $z=0$ plane, where the sample is, because of the higher strength of the field, and the waist parameter is set to $w_0=0.7\lambda$ (actual waist is $2w_0$).

### A. Tip radius

Here we assume the relative permittivity of the plasmonic tip to be $\varepsilon_t=-3.5+i0.35$, the sample radius to be $a_s=60$ nm, and the gap between the tip and sample to be $g=10$ nm (Fig. 1). The effect of the tip radius $a_t$ on the force difference $\Delta F$ between the two adopted CP polarizations is plotted in Fig. 8 for various values of the sample chirality parameter $\kappa$. It is observed that the differential force $\Delta F$ increases for tip radii larger than $a_t = 40$ nm and it is maximum for $a_t = 60$ nm in each $\kappa$ case. We have not investigated larger values for $a_t$ since the validity range of our analytical model based on dipole approximation dictates to keep the radius of the particles considerably smaller than the operational wavelength (here the largest sphere radius is such that $a_t / \lambda \leq 0.12$). Thus, in order to achieve the maximum distinction in the force between the two incident scenarios, it is better to choose $a_t = 60$ nm for the tip. Further studies could be pursued numerically or using multipole spherical harmonics.

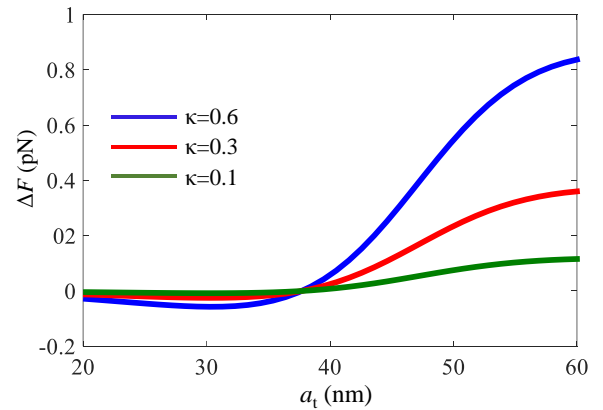

FIG. 8. Differential force $\Delta F$ for different chirality parameters of the sample as a function of the tip radius with $a_s=60$ nm, $g=10$ nm, $\varepsilon_t=-3.5+i0.35$ and $\varepsilon_s=2.5$.

### B. Relative displacement of the tip-sample system and the focus of the beam

We now investigate the robustness of the proposed method for enantio-specific detection of chiral nano-samples -by observing the sensitivity of the differential force $\Delta F$ to the position of the tip-sample interactive system with respect to the minimum waist location (maximum field strength) of the excitation beam. As we discussed earlier, we are exciting the tip-sample system with CP Gaussian beams with $w_0 = 353$ nm. First we study the effect of *lateral displacement h* of the minimum waist with respect to the tip-sample system on the differential force $\Delta F$ as shown in Fig. 9. When $h = 0$, the differential force $\Delta F$ is maximum and as $h$ increases $\Delta F$ decreases because the maximum field strength occurs along the beam axis. However, even with a displacement as large as $h = 250$ nm, samples with chirality parameter as small as $\kappa = 0.1$ could still be detectable depending on the instrument sensitiviy.

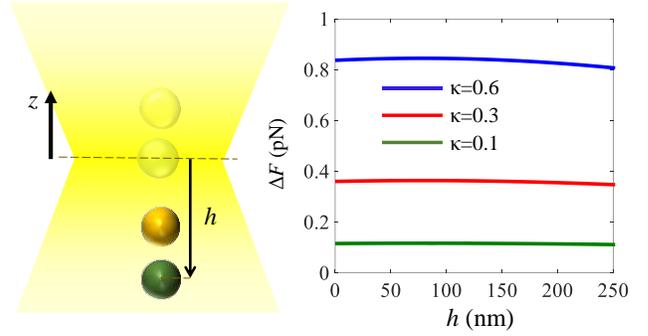

FIG. 10. Effect of longitudinal (i.e., vertical) displacement of the minimum waist with respect to the tip-sample interactive system for $a_s = a_t = 60$ nm, $g = 10$ nm, $\varepsilon_t = -3.5 + i0.35$ and $\varepsilon_s = 2.5$.

Next, we provide the results for a *longitudinal displacement h* of the minimum waist with respect to the tip-sample system as shown in Fig. 10. In this setup, when $h = 0$ the sample is at the minimum beam waist plane, whereas when $h = 130$ nm then the tip is located at the minimum waist. As noticed in Fig. 10, the longitudinal displacement of the waist with respect to the tip-sample system *does not* significantly influence the differential force $\Delta F$ since the maximum field strength is either on the sample or on the tip, and the field does not strongly change along the beam axis of a CP Gaussian beam when the waist moves in the upper direction.

### 7. CONCLUSION

We have introduced the concept of photo-induced force microscopy for enantiomer-specific detection of nanoscale chiral samples. Although we use the same excitations as applied in standard CD and ROA scenarios for detecting chirality (i.e., two CP beams with opposite handedness), the idea of using such beams in photo-induced force microscopy enables an unprecedented method for detection of chiral samples down to nanoscale resolution and sample size. That is to say, we can specify the enantiomer type of a chiral particle (with a radius as small as 25 nm) with chirality parameter as small as $\kappa = 0.04$ with nanometric resolution. We have demonstrated how an achiral plasmonic AFM tip effectively interacts with chiral nano-samples to enhance the probing force (that is in a measurable range) by providing an analytical formalism to predict the exerted force on the tip.

In a future study we plan to implement the proposed approach in an experimental set-up for the enantiomeric detection of chiral samples. We also discuss how the use of properly engineered nano-antennas with both electric and magnetic responses could maximize the measured force, hence, enabling the detection of even smaller particles or weaker chirality than the one assumed here.

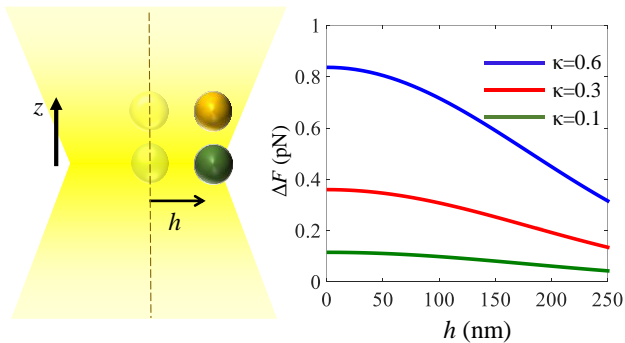

FIG. 9. Effect of lateral displacement of the tip-sample system with respect to the beam axis on differential force $\Delta F$ for $a_s = a_t = 60$ nm, $g = 10$ nm, $\varepsilon_t = -3.5 + i0.35$ and $\varepsilon_s = 2.5$.

## ACKNOWLEDGEMENTS

The authors acknowledge support from the W.M. Keck Foundation, USA.

## APPENDIX A: MIE POLARIZABILITIES FOR A CHIRAL PARTICLE

The chiral tensorial parameter $\underline{\kappa}$ is defined via the constitutive relations [38]

$$\begin{aligned} \mathbf{D} &= \varepsilon_0 \underline{\varepsilon} \cdot \mathbf{E} + i\sqrt{\varepsilon_0 \mu_0}\,\underline{\kappa} \cdot \mathbf{H} \\ \mathbf{B} &= \mu_0 \underline{\mu} \cdot \mathbf{H} - i\sqrt{\varepsilon_0 \mu_0}\,\underline{\kappa}^{\mathrm{T}} \cdot \mathbf{E}, \end{aligned} \quad (A1)$$

which provide the connection between the electric field $\mathbf{E}$, magnetic field $\mathbf{H}$, magnetic induction $\mathbf{B}$, and electric displacement vector $\mathbf{D}$. The chirality tensor $\underline{\kappa}$ provides an average measure of handedness of inclusions composing the bulk in accordance with the role of $\underline{\varepsilon}$ and $\underline{\mu}$ in the material constitutive relations (A1). Notice that the coefficient $\sqrt{\varepsilon_0 \mu_0}$ is introduced to make the chirality tensor $\underline{\kappa}$ dimensionless as $\varepsilon_0$ and $\mu_0$ are introduced to make the tensors $\underline{\varepsilon}$ and $\underline{\mu}$ dimensionless. It is shown in [38] that chirality originates from a first order spatial dispersion in material.

For a small (at the wavelength scale) sphere with radius $a$, and isotropic material parameters $\underline{\varepsilon} = \varepsilon_s \mathbf{I}$, $\underline{\mu} = \mu_s \mathbf{I}$ and $\underline{\kappa} = \kappa \mathbf{I}$ located in free space, we approximate the electromagnetic response with only dipolar terms whose electric, magnetic, and magnetoelectric polarizabilities $\alpha_s^{ee}, \alpha_s^{mm}$, and $\alpha_s^{em}$ are simply related to the material parameters via the Mie scattering as [65]

$$\alpha_s^{ee} = -6i\pi\varepsilon_0 \frac{a_1}{k^3}, \quad \alpha_s^{mm} = -6i\pi\mu_0 \frac{b_1}{k^3}, \quad \alpha_s^{em} = 6\pi\sqrt{\varepsilon_0\mu_0}\frac{c_1}{k^3}. \quad (A2)$$

Here $k$ is the free space wavenumber and coefficients $a_1$, $b_1$ and $c_1$ are given by

$$\begin{aligned} a_1 &= -\frac{1}{\Delta} i\left(1 - a_R a_L\right)\left(X_L X_R Y_2 Y_4 + U_L U_R Y_1 Y_3\right) + \\ & \quad \left(X_L U_R + X_R U_L\right)\left(a_L \mu_0 \frac{\omega Y_1 Y_4}{k} + k a_R \frac{Y_2 Y_3}{\omega \mu_0}\right) \\ b_1 &= -\frac{1}{\Delta} i\left(1 - a_R a_L\right)\left(X_L X_R Y_2 Y_4 + U_L U_R Y_1 Y_3\right) \times \\ & \quad \left(a_L \mu_0 \frac{\omega Y_2 Y_3}{k} + k a_R \frac{Y_1 Y_4}{\omega \mu_0}\right) \\ c_1 &= \frac{1}{\Delta} \Delta_1 \left(Y_2 Y_3 - Y_1 Y_4\right) \end{aligned} \quad (A3)$$

In the above equations, $\omega$ is the angular frequency and $\Delta$, $\Delta_1$, $Y_1$ to $Y_4$, $X_L$, $X_R$, $U_L$, $U_R$, $a_L$ and $a_R$ are given by

$$\begin{cases} \Delta = i\left(1 - a_R a_L\right)\left(X_L X_R Y_4^2 + U_L U_R Y_3^2\right) + \\ \quad \left(X_L U_R + X_R U_L\right)\left(a_L \mu_0 \frac{\omega}{k} + \frac{k a_R}{\omega \mu_0}\right) Y_3 Y_4 \\ \Delta_1 = X_L U_R + a_R a_L U_L X_R \end{cases} \quad (A4)$$

$$\begin{cases} Y_1 = \frac{\sin(ka)}{ka} - \frac{\cos(ka)}{ka} \\ Y_2 = \frac{1}{ka}\left(\frac{\cos(ka)}{ka} - \frac{\sin(ka)}{(ka)^2} + \sin(ka)\right) \\ Y_3 = \left(\frac{-i}{(ka)^2} - \frac{1}{ka}\right) e^{ika} \\ Y_4 = \frac{1}{ka}\left(\frac{i}{(ka)^2} + \frac{1}{ka} - i\right) e^{ika} \end{cases} \quad (A5)$$

$$\begin{cases} X_R = \frac{\sin(k_R a)}{(k_R a)^2} - \frac{\cos(k_R a)}{k_R a} \\ X_L = \frac{\sin(k_L a)}{(k_L a)^2} - \frac{\cos(k_L a)}{k_L a} \end{cases} \quad (A6)$$

$$\begin{cases} U_R = \frac{1}{k_R a}\left(\frac{\cos(k_R a)}{k_R a} - \frac{\sin(k_R a)}{(k_R a)^2} + \sin(k_R a)\right) \\ U_L = \frac{1}{k_L a}\left(\frac{\cos(k_L a)}{k_L a} - \frac{\sin(k_L a)}{(k_L a)^2} + \sin(k_L a)\right) \end{cases} \quad (A7)$$

$$\begin{cases} a_R = -i\frac{1}{\omega \varepsilon_1}\left(k_R\left(1 - \beta^2 \omega^2 \varepsilon_1 \mu_1\right) + \beta \omega^2 \varepsilon_1 \mu_1\right) \\ a_L = -i\frac{1}{\omega \varepsilon_1}\left(k_L\left(1 - \beta^2 \omega^2 \varepsilon_1 \mu_1\right) - \beta \omega^2 \varepsilon_1 \mu_1\right) \end{cases} \quad (A8)$$

In the above equations, $\varepsilon_1 = \varepsilon_0 \varepsilon_s$, $\mu_1 = \mu_0 \mu_s$ are the sphere's permittivity and permeability whereas $\beta$ is the phenomenological coefficient related to chirality parameter as

$$\beta = \frac{\kappa}{\omega\sqrt{\varepsilon_1\mu_1}}. \quad (A9)$$

Moreover $k_R$ and $k_L$ are the chiral sphere's wavenumbers for right and left hand CP waves, respectively, and given by

$$\begin{cases} k_R = \omega\sqrt{\varepsilon_1\mu_1}\left(1-\beta\omega\sqrt{\varepsilon_1\mu_1}\right)\frac{1}{1-\beta^2\omega^2\varepsilon_1\mu_1} \\ k_L = \omega\sqrt{\varepsilon_1\mu_1}\left(1+\beta\omega\sqrt{\varepsilon_1\mu_1}\right)\frac{1}{1-\beta^2\omega^2\varepsilon_1\mu_1} \end{cases} \quad (A10)$$

## APPENDIX B: DIFFERENTIAL OPTICAL FORCE EXERTED ON THE TIP

We show here the steps that lead to the approximate formula in Eq. (8) for the differential exerted force on the tip in the near field region of a chiral sample when the tip-sample system is excited by CP plane waves with opposite handedness. The expression for the time-averaged optical force exerted on the achiral tip modeled as two coexisting electric and magnetic dipoles $\mathbf{p}_t$ and $\mathbf{m}_t$ is given in Eq. (3) where $\mathbf{E}^{loc}(\mathbf{r}_t)$ and $\mathbf{H}^{loc}(\mathbf{r}_t)$ are the local electric and magnetic fields (phasors) at the tip position. They are defined as

$$\mathbf{E}^{loc}(\mathbf{r}_t) = \mathbf{E}^{inc}(\mathbf{r}_t) + \mathbf{E}_{scat}\big|_{s\to t}, \\ \mathbf{H}^{loc}(\mathbf{r}_t) = \mathbf{H}^{inc}(\mathbf{r}_t) + \mathbf{H}_{scat}\big|_{s\to t}, \quad (B1)$$

in which $\mathbf{E}^{inc}$ and $\mathbf{H}^{inc}$ are the incident (i.e., external) electric and magnetic fields evaluated at the tip location. Furthermore, $\mathbf{E}_{scat}\big|_{s\to t}$ and $\mathbf{H}_{scat}\big|_{s\to t}$ are the scattered fields generated by the sample at the tip position. Despite all the calculations in the paper use all the dynamic terms of the Green's function [39], here we approximate the scattered fields by an electric $\mathbf{p}$ and a magnetic $\mathbf{m}$ dipole in the near field region by retaining only the stronger term [66]

$$\mathbf{E}_{scat} \approx \frac{e^{ikr}}{4\pi\varepsilon_0 r^3}\left[3\hat{\mathbf{r}}(\hat{\mathbf{r}}\cdot\mathbf{p})-\mathbf{p}\right], \\ \mathbf{H}_{scat} \approx \frac{e^{ikr}}{4\pi\mu_0 r^3}\left[3\hat{\mathbf{r}}(\hat{\mathbf{r}}\cdot\mathbf{m})-\mathbf{m}\right], \quad (B2)$$

in which $\hat{\mathbf{r}} = \mathbf{r}/r$ is the unit vector of the radial direction in spherical coordinates centered at the source location, $\mathbf{r}$ is the vector from the source to the observation point and $k$ is the ambient wavenumber. Consequently, the scattered fields due to sample at the tip position are simply rewritten as

$$\mathbf{E}_{scat}\big|_{s\to t} = -\frac{G}{\varepsilon_0}\left[p_{s,x}\hat{\mathbf{x}} + p_{s,y}\hat{\mathbf{y}} - 2p_{s,z}\hat{\mathbf{z}}\right], \\ \mathbf{H}_{scat}\big|_{s\to t} = -\frac{G}{\mu_0}\left[m_{s,x}\hat{\mathbf{x}} + m_{s,y}\hat{\mathbf{y}} - 2m_{s,z}\hat{\mathbf{z}}\right], \quad (B3)$$

in which

$$G = \frac{e^{ik|z_t-z_s|}}{4\pi|z_t-z_s|^3}. \quad (B4)$$

Note that subscripts $x$, $y$ and $z$ are representing the related components in the Cartesian coordinates, $z_t$ and $z_s$ are the tip and sample positions, respectively.

The electric and magnetic dipole moments of the sample, assumed to have isotropic polarizability, are calculated as

$$\mathbf{p}_s = \alpha_s^{ee}\mathbf{E}^{loc}(z_s) + \alpha_s^{em}\mathbf{H}^{loc}(z_s), \\ \mathbf{m}_s = \alpha_s^{me}\mathbf{E}^{loc}(z_s) + \alpha_s^{mm}\mathbf{H}^{loc}(z_s), \quad (B5)$$

where the corresponding sample polarizabilities $\alpha_s$ are defined in the manuscript. Moreover, the magnetic dipole moment associated to a current density in a given volume is defined as $\mathbf{m} = \frac{1}{2}\mu_0\int dv\, \mathbf{r}\times\mathbf{J}$, with $\mathbf{J}$ and $\mathbf{r}$ being the volumetric current and the position vector in the same volume, respectively. Next, by using Eqs. (B3) and (B5), the local fields at the tip location in Eq. (B1) are found to be

$$\begin{bmatrix} E_x^{loc}(z_t) \\ E_y^{loc}(z_t) \\ E_z^{loc}(z_t) \end{bmatrix} = \begin{bmatrix} E_x^{inc}(z_t) \\ E_y^{inc}(z_t) \\ E_z^{inc}(z_t) \end{bmatrix} - \frac{G}{\varepsilon_0}\begin{bmatrix} \alpha_s^{ee}E_x^{loc}(z_s) + \alpha_s^{em}H_x^{loc}(z_s) \\ \alpha_s^{ee}E_y^{loc}(z_s) + \alpha_s^{em}H_y^{loc}(z_s) \\ -2\alpha_s^{ee}E_z^{loc}(z_s) - 2\alpha_s^{em}H_z^{loc}(z_s) \end{bmatrix}, \quad (B6)$$

$$\begin{bmatrix} H_x^{loc}(z_t) \\ H_y^{loc}(z_t) \\ H_z^{loc}(z_t) \end{bmatrix} = \begin{bmatrix} H_x^{inc}(z_t) \\ H_y^{inc}(z_t) \\ H_z^{inc}(z_t) \end{bmatrix} - \frac{G}{\mu_0}\begin{bmatrix} \alpha_s^{me}E_x^{loc}(z_s) + \alpha_s^{mm}H_x^{loc}(z_s) \\ \alpha_s^{me}E_y^{loc}(z_s) + \alpha_s^{mm}H_y^{loc}(z_s) \\ -2\alpha_s^{me}E_z^{loc}(z_s) - 2\alpha_s^{mm}H_z^{loc}(z_s) \end{bmatrix}, \quad (B7)$$

where the local electric and magnetic fields at the sample position can be obtained by

$$\mathbf{E}^{\text{loc}}(z_s) = \mathbf{E}^{\text{inc}}(z_s) + \mathbf{E}_{\text{scat}}|_{t \to s},$$
$$\mathbf{H}^{\text{loc}}(z_s) = \mathbf{H}^{\text{inc}}(z_s) + \mathbf{H}_{\text{scat}}|_{t \to s}, \quad (B8)$$

in which $\mathbf{E}_{\text{scat}}|_{t \to s}$ and $\mathbf{H}_{\text{scat}}|_{t \to s}$ are the electric and magnetic fields scattered by the tip at the sample position and similar to Eq. (B3) are calculated by

$$\mathbf{E}_{\text{scat}}|_{t \to s} = -\frac{G}{\varepsilon_0}\left[ p_{t,x}\hat{\mathbf{x}} + p_{t,y}\hat{\mathbf{y}} - 2p_{t,z}\hat{\mathbf{z}} \right],$$
$$\mathbf{H}_{\text{scat}}|_{t \to s} = -\frac{G}{\mu_0}\left[ m_{t,x}\hat{\mathbf{x}} + m_{t,y}\hat{\mathbf{y}} - 2m_{t,z}\hat{\mathbf{z}} \right]. \quad (B9)$$

Since the tip is achiral, the electric and magnetic dipole moments of the tip read

$$\mathbf{p}_t = \alpha_t^{ee}\mathbf{E}^{\text{loc}}(z_t),$$
$$\mathbf{m}_t = \alpha_t^{mm}\mathbf{H}^{\text{loc}}(z_t). \quad (B10)$$

Now, combining Eqs. (B8)-(B10) and considering that the incident CP plane waves (note that in the numerical examples in the paper we assumed structures to be illuminated by CP Gaussian beams whereas here, for the sake of simplicity, we assume plane waves) lack $z$-polarized field components, Eqs. (B6) and (B7) read

$$\begin{bmatrix} E_x^{\text{loc}}(z_t) \\ E_y^{\text{loc}}(z_t) \end{bmatrix} = \begin{bmatrix} E_x^{\text{inc}}(z_t) \\ E_y^{\text{inc}}(z_t) \end{bmatrix} - \frac{G}{\varepsilon_0} \times$$
$$\begin{bmatrix} \alpha_s^{ee}\left(E_x^{\text{inc}}(z_t) - \frac{G}{\varepsilon_0}\alpha_t^{ee}E_x^{\text{loc}}(z_t)\right) + \alpha_s^{em}H_x^{\text{inc}}(z_t) \\ \alpha_s^{ee}\left(E_y^{\text{inc}}(z_t) - \frac{G}{\varepsilon_0}\alpha_t^{ee}E_y^{\text{loc}}(z_t)\right) + \alpha_s^{em}H_y^{\text{inc}}(z_t) \end{bmatrix}, \quad (B11)$$

and

$$\begin{bmatrix} H_x^{\text{loc}}(z_t) \\ H_y^{\text{loc}}(z_t) \end{bmatrix} = \begin{bmatrix} H_x^{\text{inc}}(z_t) \\ H_y^{\text{inc}}(z_t) \end{bmatrix} - \frac{G}{\mu_0} \times$$
$$\begin{bmatrix} \alpha_s^{me}\left(E_x^{\text{inc}}(z_t) - \frac{G}{\varepsilon_0}\alpha_t^{ee}E_x^{\text{loc}}(z_t)\right) + \alpha_s^{mm}H_x^{\text{inc}}(z_t) \\ \alpha_s^{me}\left(E_y^{\text{inc}}(z_t) - \frac{G}{\varepsilon_0}\alpha_t^{ee}E_y^{\text{loc}}(z_t)\right) + \alpha_s^{mm}H_y^{\text{inc}}(z_t) \end{bmatrix}. \quad (B12)$$

The $z$-component of the local fields at the tip location are not shown above since it is vanishing. In obtaining the above equations, we have used the assumption that at resonance of the chiral sample the following approximation holds $\alpha_s^{ee}\alpha_s^{mm} \approx \alpha_s^{me}\alpha_s^{em}$ [67–69]. This is a good approximation for most dipole scatterers close to their resonance and when both the electric and magnetic dipole response originate from the same equation of motion for charge. Next, by simplifying Eqs. (B11) and (B12), for the $x$-component of the electric and magnetic fields we get

$$E_x^{\text{loc}}(z_t) = \frac{1}{1 - G^2 \frac{\alpha_t^{ee}}{\varepsilon_0}\frac{\alpha_s^{ee}}{\varepsilon_0}} \times$$
$$\left[\left(1 - G\frac{\alpha_s^{ee}}{\varepsilon_0}\right)E_x^{\text{inc}}(z_t) - \frac{\alpha_s^{em}}{\varepsilon_0}H_x^{\text{inc}}(z_t)\right], \quad (B13)$$

and

$$H_x^{\text{loc}}(z_t) = \left(-G\frac{\alpha_s^{me}}{\mu_0} + \frac{\left(1 - G\frac{\alpha_s^{ee}}{\varepsilon_0}\right)G^2}{1 - G^2\frac{\alpha_t^{ee}}{\varepsilon_0}\frac{\alpha_s^{ee}}{\varepsilon_0}}\frac{\alpha_t^{ee}}{\varepsilon_0}\frac{\alpha_s^{me}}{\mu_0}\right)E_x^{\text{inc}}(z_t) +$$
$$\left(1 - G\frac{\alpha_s^{mm}}{\mu_0} - \frac{G^3}{1 - G^2\frac{\alpha_t^{ee}}{\varepsilon_0}\frac{\alpha_s^{ee}}{\varepsilon_0}}\frac{\alpha_t^{ee}}{\varepsilon_0}\frac{\alpha_s^{me}}{\mu_0}\frac{\alpha_s^{em}}{\varepsilon_0}\right)H_x^{\text{inc}}(z_t). \quad (B14)$$

Similar expressions can be obtained for the $y$-component of the fields. The incident fields for the CP waves propagating along the $z$-direction are

$$\mathbf{E}^{\text{inc}} = |\mathbf{E}_0|\frac{(\hat{\mathbf{x}} \pm i\hat{\mathbf{y}})}{\sqrt{2}}e^{ikz}$$
$$\mathbf{H}^{\text{inc}} = \frac{|\mathbf{E}_0|}{\eta_0}\frac{(\mp i\hat{\mathbf{x}} + \hat{\mathbf{y}})}{\sqrt{2}}e^{ikz} \quad (B15)$$

in which +/− signs represent right (upper sign) and left (lower sign) handed CP waves, with electric field magnitude $|\mathbf{E}_0|$. By replacing these incident fields in Eqs. (B13) and (B14) we find the local fields of the tip and by inserting the values of the local fields in Eq. (B10) we obtain the dipole moments. Then, we use Eq. (3) to find the exerted force on the tip. As was emphasized in the paper, we neglect the magnetic dipole moment of the tip in obtaining our approximate formula, thus based on Eq. (3) the differential force is calculated as

$$\Delta F = |\mathbf{E}_0|^2 \text{Im}\left\{\frac{\alpha_t^{ee}}{\sqrt{\varepsilon_0\mu_0}d}\left(\frac{Num}{Den}\right)\right\}, \quad (B16)$$

with

$$Num = \Lambda\left[1-(1-ikd)\frac{\alpha_s^{ee}}{4\pi d^3}\right]\frac{\left(\alpha_s^{em}\right)^*}{4\pi d^3} +$$

$$\left(\Lambda - 3\frac{\left(\alpha_t^{ee}\right)^*}{2\pi d^3}\right)\left[(1-ikd)\frac{\left(\alpha_s^{ee}\right)^*}{4\pi d^3}\right]\frac{\alpha_s^{em}}{4\pi d^3} \quad (B17)$$

$$Den = \left[1-\frac{\left(\alpha_t^{ee}\right)^*}{4\pi d^3}\frac{\left(\alpha_s^{ee}\right)^*}{4\pi d^3}\right]^2\left[1-\frac{\alpha_t^{ee}}{4\pi d^3}\frac{\alpha_s^{ee}}{4\pi d^3}\right]$$

where

$$\Lambda = (3+2ikd)+(3+4ikd)\frac{\left(\alpha_t^{ee}\right)^*\left(\alpha_s^{ee}\right)^*}{4\pi d^3\;4\pi d^3}, \quad (B18)$$

Note that in the calculation of Eq. (B16) we take the gradients of electric and magnetic field vectors $\mathbf{E}^{loc}$ and $\mathbf{H}^{loc}$. In Cartesian coordinates, the gradient of a representative vector $\mathbf{A}$ is defined as

$$\nabla\mathbf{A} = \begin{bmatrix}\partial_x A_x & \partial_y A_x & \partial_z A_x \\ \partial_x A_y & \partial_y A_y & \partial_z A_y \\ \partial_x A_z & \partial_y A_z & \partial_z A_z\end{bmatrix}.$$

Next, considering the quasi-static limit by assuming $kd \to 0$ and neglecting the terms that contain polarizability power orders higher than the second, Eq. (B16) simplifies to Eq. (8). When either the tip or the sample is lossless Eq. (8) is further simplified to

$$\Delta F \approx -\frac{3|\mathbf{E}_0|^2}{4\pi d^4\sqrt{\varepsilon_0\mu_0}}\mathrm{Re}\{\alpha_t^{ee}\}\mathrm{Im}\{\alpha_s^{em}\}. \quad (B19)$$

We recall that $\alpha_s^{em} = -\alpha_s^{me}$, and that under the assumption of negligible sample losses $\alpha_s^{me}$ is purely imaginary [38].

## APPENDIX C: EQUAIVALENT REPRESENTATION OF EFFECTIVE POLARIZABILITY OF THE TIP

In section 4 of the paper we explained how optical activity is transferred from a chiral sample to an achiral tip thanks to their near field coupling. Here we show the details that lead to Eq. (13). We first insert Eqs. (11) and (12) into Eq. (10). Then, based on Eq. (7), the induced force on the tip in the *z*-direction reads

$$F_z^\pm = \frac{|\mathbf{E}_0|^2}{2|\alpha_t^{ee}|^2}\mathrm{Re}\left\{\alpha_t^{ee}\left(\hat{\alpha}^\pm\frac{\partial\left(\hat{\alpha}^\pm\right)^*}{\partial z}-ik|\hat{\alpha}^\pm|^2\right)\right\}, \quad (C1)$$

Where the + and − signs represent right and left handed CP waves, respectively. In Eq. (C1) we have defined the effective polarizability parameter $\hat{\alpha}^\pm$ for the right/left CP wave as

$$\hat{\alpha}^\pm = \hat{\alpha}_t^{ee} \mp i\hat{\alpha}_t^{em}. \quad (C2)$$

where $\hat{\alpha}_t^{ee}$ and $\hat{\alpha}_t^{em}$ are defined in Eqs. (11) and (12). Therefore, the differential force is calculated as

$$\Delta F = F_z^{RCP} - F_z^{LCP} =$$

$$\frac{|\mathbf{E}_0|^2}{2|\alpha_t^{ee}|^2}\mathrm{Re}\left\{\alpha_t^{ee}\left(\hat{\alpha}^+\frac{\partial\left(\hat{\alpha}^+\right)^*}{\partial z}-\hat{\alpha}^-\frac{\partial\left(\hat{\alpha}^-\right)^*}{\partial z}\right)\right\}\Bigg|_{z=z_t}.$$

(C3)

The derivative is taken with respect to the observation *z*-coordinate (i.e., the tip location). Now, inserting Eq. (C2) into Eq. (C3) leads to Eq. (13).